\documentclass[12pt,preprint]{aastex}

\def\gsim{\ \raise 3pt \hbox{$>$} \kern -8.5pt \raise -2pt \hbox{$\sim$}\ }
\def\lsim{\ \raise 3pt \hbox{$<$} \kern -8.5pt \raise -2pt \hbox{$\sim$}\ }

\begin{document}
\title{Spatial Evidence for Transition Radiation in a Solar Radio Burst}
\author{Gelu M. Nita\altaffilmark{1}, Dale E. Gary\altaffilmark{1}, and Gregory D. Fleishman\altaffilmark{1, 2}}
\affil{$^1$New Jersey Institute of Technology, Newark, NJ 07102,
$^2$National Radio Astronomy Observatory, Charlottesville 22903, VA,
USA}

\begin{abstract}
Microturbulence, i.e. enhanced fluctuations of plasma density,
electric and magnetic fields, is of great interest in
astrophysical plasmas, but occurs on spatial scales far too small
to resolve by remote sensing, e.g., at $\sim$1-100 cm in the solar
corona.  This paper reports spatially resolved observations that
offer strong support for the presence in solar flares of a
suspected radio emission mechanism, resonant transition radiation,
which is tightly coupled to the level of microturbulence and
provides direct diagnostics of the existence and level of
fluctuations on decimeter spatial scales. Although the level of
the microturbulence derived from the radio data is not
particularly high, $\left<\Delta n^2 \right>/n^2 \sim 10^{-5}$, it
is large enough to affect the charged particle diffusion and give
rise to effective stochastic acceleration. This finding has
exceptionally broad astrophysical implications since modern
sophisticated numerical models predict generation of much stronger
turbulence in relativistic objects, e.g., in gamma-ray burst
sources.
\end{abstract}


\section{Introduction}

Microturbulence in cosmic sources  governs the dynamics of energy
release and dissipation in astrophysical and geospace plasmas, the
formation of collisionless shock waves and current sheets, and is
a key ingredient in stochastic acceleration \citep{c1,c2} and
enhanced diffusion \citep{DT_1966,c3,c4} of nonthermal particles.
The microturbulence may also  affect the electromagnetic emission
produced by fast particles, giving rise to Transition Radiation
(TR), which was proposed nearly 60 years ago by two Nobel Prize
winning physicists, \citet{c5}. TR in its original form \citep{c5}
results from a variation in phase speed of wave propagation at
transition boundaries. The theory of TR has seen wide application
in the laboratory \citep{c6} and in cosmic ray detectors
\citep{c7,c8}, although no naturally occurring radiation had been
confirmed as TR.

In the astrophysical context, TR must arise whenever nonthermal
charged particles pass near or through small-scale inhomogeneities
such as wave turbulence or dust grains.  However, it was thought
to be weak, and perhaps unobservable \citep{c9b, c9a, c9}, until
\citet{c10} showed that its intensity can be greatly enhanced due
to plasma resonance at frequencies just above the local plasma
frequency. Spatially and spectrally resolved observations of this
resonant transition radiation (RTR), if present, can provide
quantitative diagnostics of plasma density, and of the level of
microturbulence in the flaring region.

A number of recent publications, based mainly on studies of
individual events,  indicate that RTR may be produced in solar
radio bursts \citep{c10a,c10b,c10c,c10d,Bogod_Yasnov_2005}. Most
recently, we have described the observational characteristics
expected for RTR in the case of solar bursts \citep{c11}, and
found that the correlations and associations predicted for total
power data are indeed present in the decimetric ($\sim$1-3 GHz)
components of a statistical sample of two-component solar
continuum radio bursts. However, interpretations based on
non-imaging  data remain indirect (and, thus, ambiguous) until
they can be combined with direct imaging evidence from
multi-wavelength spatially resolved observations, which  were
missing in the previous studies.

This report presents comprehensive (radio, optical, and soft
X-ray) spatially resolved observations for one of the
RTR-candidate bursts.  As we describe below,  these observations
provide primarily three new confirmations: (1) the RTR and
gyroemission sources are co-spatial, (2) the RTR component is
associated with a region of high density, and (3) the RTR emission
is $o$-mode polarized. Together with the already demanding
spectral and polarization correlations found previously
\citep{c11}, these new observations provide further strong
evidence in favor of RTR.

\section{Theoretical Expectations}

The two spectral components of such RTR candidate bursts (one at
centimeter wavelengths due to the usual gyrosynchrotron (GS) mechanism,
and one at decimeter wavelengths suspected as RTR), must be
\emph{co-spatial} to allow an unambiguous RTR interpretation. An
alternative explanation of such two-component bursts is that both
spectral components are produced by the same (GS)
emission mechanism (with different parameter combinations in the
two components), but if the low and high frequencies come from the
same source location this should merely broaden the spectrum.
Having truly separate spectral components requires either
completely different source locations, or different mechanisms, or
both. Distinct spectral components having the \emph{same source
location} is a strong indicator that each component is produced by
\emph{a different emission mechanism}. Therefore, direct
observation of the spatial relationship between the spectral
components of RTR candidate bursts is the key evidence needed to
conclude the emission mechanism producing the decimetric spectral
component.

The theory of RTR in the astrophysical context is discussed in
detail in a recent review paper \citep{c12}. RTR arises as fast
particles move through a plasma with small-scale variations (as
short as the wavelength of the emitted wave) of the refractive
index.  Such variations may be provided by microturbulence-induced
inhomogeneities of the plasma density or magnetic field.

In the case of solar bursts, the main properties of this emission
mechanism that can be checked against observations are: The
emission (1) originates in a dense plasma, $f_{pe} \gg f_{Be}$,
where $f_{pe}$ and $f_{Be}$ are the electron plasma- and
gyro-frequencies; (2) has a relatively low peak frequency in the
decimetric range, and so appears as a low-frequency component
relative to the associated GS spectrum; (3) is co-spatial with or
adjacent to the associated GS source; (4) varies with a time scale
comparable to the accompanying GS emission (assuming a constant or
slowly varying level of the necessary microturbulence); (5) is
typically strongly polarized in the ordinary mode ($o$-mode),
since the extraordinary mode ($x$-mode) is evanescent, as for any
radiation produced at the plasma frequency in a magnetized plasma;
(6) is produced typically by the lower-energy end of the same
nonthermal electron distribution that produces the GS emission,
with  the \emph{emissivity proportional to the instantaneous total
number} of the low-energy electrons in the source \emph{at all
times during the burst} (in contrast to plasma emission, whose
highly nonlinear emissivity is largely decoupled from electron
number even though it may for a time display a similar
proportionality); (7) has a high-frequency spectral slope that
does not correlate with the spectral index of fast electrons (in
contrast to GS radiation, which does).

\section{Data Analysis}

Figure 1 presents the dynamic spectrum of the 2001 April 06 solar
radio burst in intensity and circular polarization, observed with
the Owens Valley Solar Array \citep{c13} (OVSA). This event is one
of many observed with OVSA whose spectral behavior matches the
expectations for RTR \citep{c11}, but is the first for which
detailed spatial comparison has been made. The RTR occurs at a
restricted range of time and frequency shown by the bright red
region in the bottom panel, which represents highly right hand
circularly polarized (RCP) emission.  The results presented in
Figures 2 and 3 confirm the expected spatial association of the
RTR radio source with (i) the accompanying GS source, (ii) an
unusually dense soft X-ray loop, and (iii) the underlying magnetic
field structure, and hence offer further support for its
interpretation as RTR emission. Comparing the required
observational characteristics in the order 1--7 presented above,
we find:
\begin{enumerate}
\item Both the RTR (2 GHz) and GS (7.4 GHz) sources arise in or
near an unusually dense loop. The electron temperature inferred
from SXT data (Fig. 3), averaged over the pixels lying inside the
85\% 2 GHz RCP contour, is $2\times10^7$ K, while the average
emission measure corresponding to one pixel ($2.5\times2.5$
arcsec) is $5.6\times10^{48}$ cm$^{-3}$. Assuming a line of sight
length of $\sim$25 arcsec,  the projected loop width, we obtain an
estimate for the plasma density in the region as $3\times10^{11}$
cm$^{-3}$. This value directly confirms the existence of a high
plasma density in the flaring region, as suggested by the Razin
effect diagnosis we employed previously \citep{c11}. The RTR peak
frequency of 2 GHz implies, from the electron plasma frequency
$f_{pe}=9\times10^3\sqrt{n}$ Hz, an electron density of
$5\times10^{10}$ cm$^{-3}$, compared with $3\times10^{11}$
cm$^{-3}$ derived above for the underlying soft X-ray loop. The
X-ray-derived density demonstrates the presence of high densities
in the region, while the lower radio-derived density is expected
since the 2 GHz radio emission will come primarily from overlying,
less-dense regions due to significant free-free absorption in the
higher-density regions.

\item As seen in Fig. 1, the RTR forms a distinct, low-frequency
spectral component relative to the higher-frequency
GS component.

\item Figs. 2 and 3 show that the RTR and GS sources are
co-spatial. As already emphasized, this co-spatiality is highly
conclusive in favor of RTR, since separate spectral components (of
multi-component bursts) typically come from distinct locations
\citep{c14,c15}.

\item Both spectral components are smooth in time and frequency, with
comparable time scales, the main difference being that the GS
component is delayed with respect to the RTR component (see Figs. 1
and 4). Note also in Fig. 4 the similarity of high-energy HXR with
the GS (7.4 GHz) component, and low-energy HXR with the RTR (2 GHz)
component, which we discuss in more detail in item 6, below.

\item Figs. 1 and 2 show that the RTR emission is strongly
polarized in the sense of the $o$-mode, as required, while the
GS emission is $x$-mode. The radio maps at 7.4 GHz in
Fig. 2 (filled contours) reveal RCP (red) overlying positive
(white) magnetic polarity and LCP (blue) overlying negative
(black) polarity, located on opposite sides of the neutral line.
This clearly shows a relatively high degree of $x$-mode
polarization of both 7.4 GHz radio sources. At 2 GHz (unfilled
contours), exactly the opposite spatial correspondence is seen,
with RCP (red) overlying negative magnetic polarity and (the much
weaker) LCP (blue) slightly shifted toward positive polarity. This
clearly shows a high degree of $o$-mode polarization for the RTR
spectral component.

\item Indirect statistical evidence for the RTR component being
due to low energy electrons was obtained from spectral
correlations \citep{c11}. \footnote{Note that in most incoherent
emission mechanisms, the spatially resolved brightness temperature
provides a lower limit to the energy of emitting electrons. The
brightness temperature of the 2 GHz RCP source in Fig. 2 reaches
$2.5\times10^9$ K, which for the typical incoherent mechanism
would correspond to a particle energy of about 220 keV (indeed
lower than the energy of the synchrotron emitting fast electrons
specified below). However, for the RTR case this argument is
inconclusive since the brightness temperature of RTR depends on
effective energies of both fast electrons and nonthermal density
fluctuations, rather than of fast electrons only.} A more reliable
estimate of the energy of the fast electrons involved comes from a
comparison of the radio and hard X-ray light-curves in Figure 4.
We first note the similarity of the RTR light curve and the 41-47
keV hard X-ray light curve. As shown by \citet{c17}, hard X-rays
are due to electrons of energy 2-3 times higher than the photon
energy, so that 41-47 keV HXR correspond to $\sim 100-150$ keV
electrons. In contrast, the GS light curve at 7.4 GHz displays a
poor correlation with 41-47 keV HXR, but an excellent correlation
with the higher energy HXR light curve, at 128-157 keV, produced
by the electrons of $\sim 250-450$ keV. This is consistent with
the well known result that GS emission comes from electrons of
energy typically $> 300$ keV \citep{c16}. The similarity of the
shape and timing of the 128-157 keV HXR and 7.4 GHz light curves,
and those of the 41-47 keV HXR and 2 GHz light curves, is
consistent with their being due to electrons of energies $\gsim
300$ keV and $\lsim 150$ keV, respectively. It is reasonable to
conclude that the RTR and GS emission, being essentially
co-spatial, are from different parts of a single electron energy
distribution.

\item As reported in an earlier paper \citep[fig.7]{c11} the
high-frequency slopes of the RTR and GS spectra for this event are
uncorrelated, which provides an independent confirmation that the
low-frequency component is not simply a low-frequency GS source.

\end{enumerate}

\section{Discussion}

The above characteristics rule out standard GS emission for the
low-frequency spectral component, while they are expected and
agree fully with RTR. An alternative model that might account for
the presence of a co-spatial, yet distinct dm-continuum spectral
component---quasi-stationary plasma emission due to a marginally
stable regime of a loss-cone instability---is much more difficult
to eliminate, or even distinguish from RTR. Indeed, properties 1,
2, 6, 7 are typical also for plasma emission, and properties 3, 4,
5, while not required for plasma emission, are not inconsistent
with it. We believe that the key evidence distinguishing RTR from
plasma emission is the strict proportionality between the radio
flux and the number of emitting electrons on all time scales, as
suggested by the agreement between the RTR time profile and the
low-energy hard X-ray light curve of Fig. 4. This proportionality,
based on the spectral properties of the dm bursts, was found in
all of the bursts studied by \cite{c11}.  We note, however, that a
temporal resolution better than the 4 s we have available will be
needed to check this property down to millisecond time scales.

Nevertheless, we looked for further evidence favoring the plasma
emission interpretation of the smooth dm component and conclude
that this model (even though not firmly eliminated) is not
supported by the data. For example, the high degree of $o$-mode
polarization of the dm continuum implies fundamental rather than
harmonic plasma emission, although the latter is typically much
easier to generate in the coronal plasma. However, spectra in this
burst and in the other bursts studied by \cite{c11}, at no time
show any hint of a second harmonic spectral feature. Furthermore,
quasi-steady plasma emission requires a significant loss-cone
anisotropy, which in turn gives rise to a widely observed loop-top
peak brightness for the optically thin GS radio emission
\citep{Meln_etal_2002}. In contrast, the 7.4 GHz source displays a
clear separation into $x$-polarized kernels (corresponding to leg
or foot-point sources, rather than a loop-top source), thus, any
pitch-angle anisotropy is at best very modest. This conclusion is
also supported by the statistical evidence found in \citep{c11} in
favor of more isotropic (than on average) distributions of the
fast electrons in the RTR-producing bursts. Therefore, all the
properties specific for RTR and those common for both RTR and
plasma emission are observed, while no specific property expected
solely for the plasma emission is seen, which leads us to favor
RTR.

We have presented ample evidence that the decimetric component of
the 2001 April 06 radio burst near 19:23 UT is produced by the RTR
mechanism. Since this event is one among a set of other events
with similar, unique characteristics, the evidence presented here
supports the conclusions made by \citet{c11}, based on total power
data for a statistical sample of the bursts candidates, that these
bursts are due to RTR.

The importance of this result is several-fold. First, it
strengthens the case for RTR as another incoherent continuum
emission mechanism in astrophysical plasmas, among only a small
number of others: gyrosynchrotron/synchrotron emission,
bremsstrahlung, and inverse Compton emission. Second, there are a
few types of solar radio continuum, e.g., type I and type IV m/dm,
which are conventionally ascribed to plasma emission. We point out
that this interpretation has never been quantitatively proved, and
RTR represents a plausible alternative to the current
interpretation, which we believe calls for revisiting the issue of
the origin of non-GS solar radio continua. Third, with new radio
facilities in development that are capable of simultaneous spatial
and spectral measurements of solar bursts (e.g. Expanded VLA
\citep{c18} and Frequency Agile Solar Radiotelescope (FASR)
\citep{c19}), RTR can be routinely recognized and used as a
diagnostic of the plasma density, the low-energy part of the
electron energy distribution, and of the presence and quantitative
level of microturbulence.  In this event, for example, the level
of inhomogeneities derived from the RTR flux, described by Eq.
(403) in \citet{c12}, is $\left<\Delta n^2 \right>/n^2 \sim
10^{-5}$.  Thus, RTR may provide a sensitive tool for measuring
this elusive but important quantity.

\acknowledgments We acknowledge NSF grant AST-0307670 and NASA
grant NAG5-11875 to NJIT. The NRAO is a facility of the NSF
operated under cooperative agreement by Associated Universities,
Inc. We gratefully acknowledge the help of J. Qiu in providing the
X-ray and MDI data.

\clearpage

\begin{figure}
\plotone{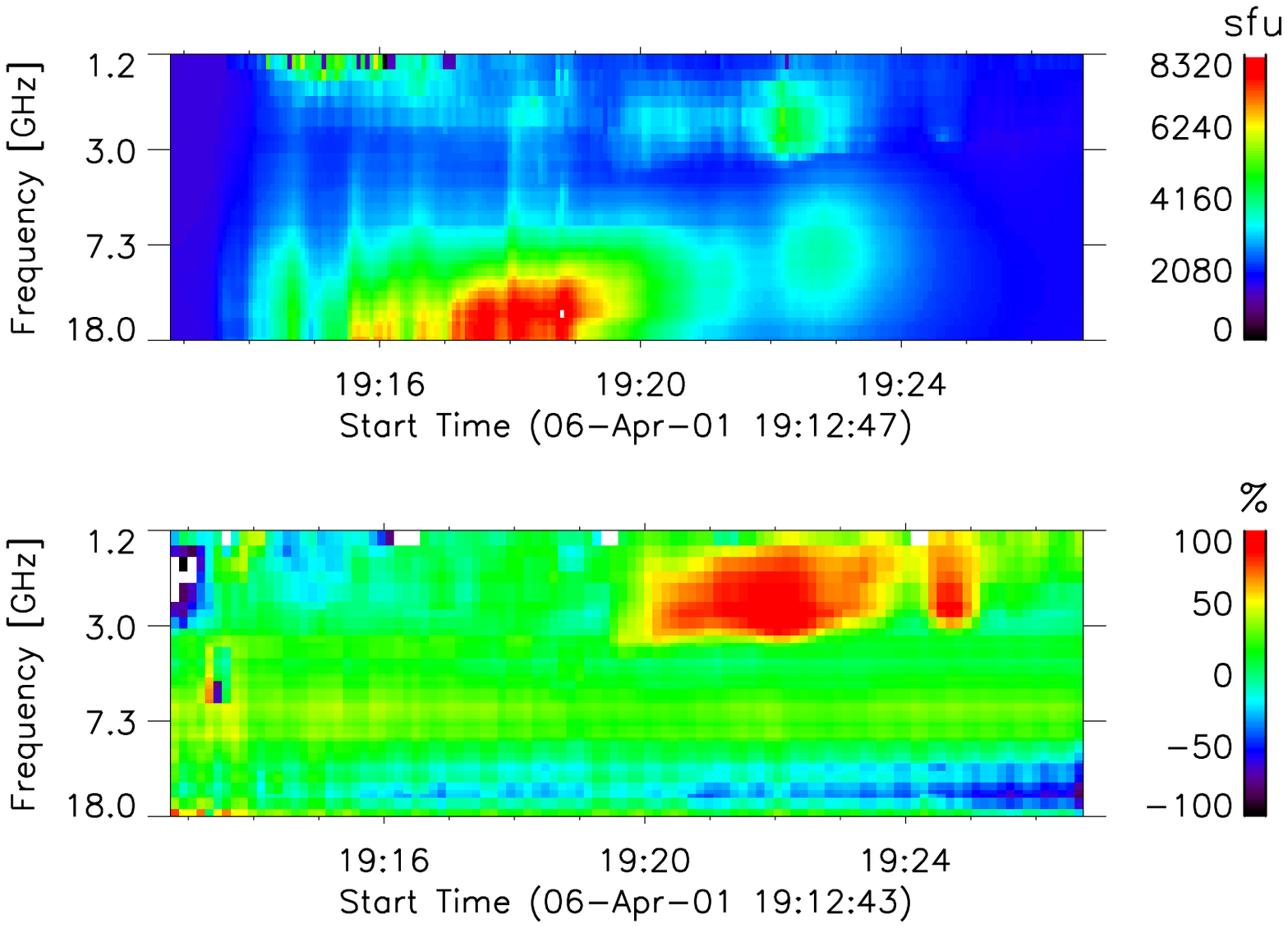}
\caption{2001 April 06 after 19:12 UT. Upper panel:
Total power dynamic spectrum recorded by OVSA with 4 s time
resolution at 40 frequencies in the [1.2--18] GHz range. Lower
panel: Dynamic spectrum of circular polarization with 8 s time
resolution at the same frequencies as in the upper panel.  The
period of RTR is the highly polarized (red) emission in the lower
panel.  Two spectral components are visible in the upper panel
during this time: the low frequency RTR component, which peaks at
19:22:11 UT (3700 sfu at 2 GHz), and the delayed high frequency GS
component, which peaks at 19:22:51 UT (2300 sfu at 7.4 GHz).}
\end{figure}

\clearpage

\begin{figure}
\epsscale{0.75} \plotone{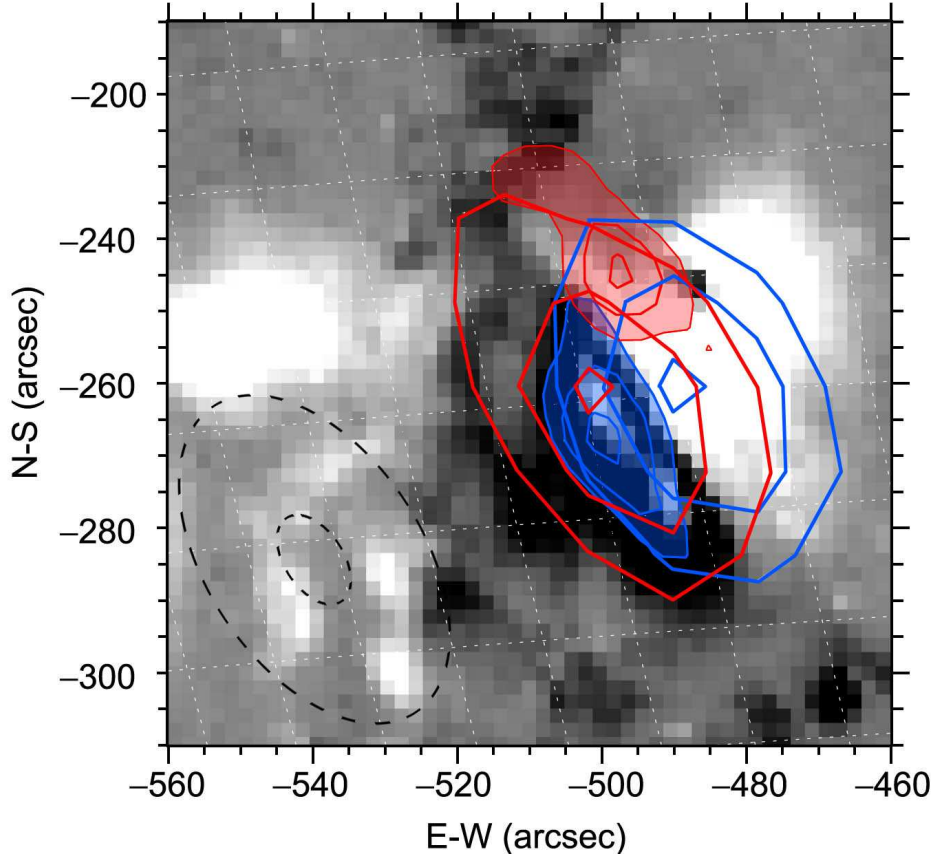} \caption{\small 2001 April 06,
OVSA radio maps (19:22:03 UT). The RCP (red contours) and LCP
(blue contours) at 2 GHz (unfilled contours) and 7.4 GHz (filled
contours) are overlaid on the SOHO \citep{c22} MDI magnetogram
(19:22:02 UT). The dashed ovals represent the half power OVSA beam
at the two selected frequencies. The radio contours are scaled
separately for each frequency and polarization, and only 3 are
shown for clarity, representing 55, 75, and 95\% of the maximum
intensity. The maximum brightness temperatures are 2500 MK (2 GHz
RCP), 770 MK (2 GHz LCP), 880 MK (7.4 GHz RCP) and 600 MK (7.4 GHz
LCP). Small islands of apparent magnetic field sign reversal in
regions of both polarities are an instrumental artifact
\citep{c23} and not real. Within the instrumental resolution (see
the corresponding beam size), the 2 GHz RCP source (red, unfilled
contour) is co-located with the 7.4 GHz LCP source (blue, filled
contour) in the negative magnetic field region. We conclude that
both low and high frequency emissions are likely produced by the
same population of electrons travelling along the same magnetic
loop. Remarkably, for both frequencies, the \emph{intrinsic}
degrees of polarization implied by the radio maps are noticeably
larger than those suggested by the unresolved polarization
spectrum presented in the lower panel of Fig. 1.}
\end{figure}

\clearpage

\begin{figure}
\plotone{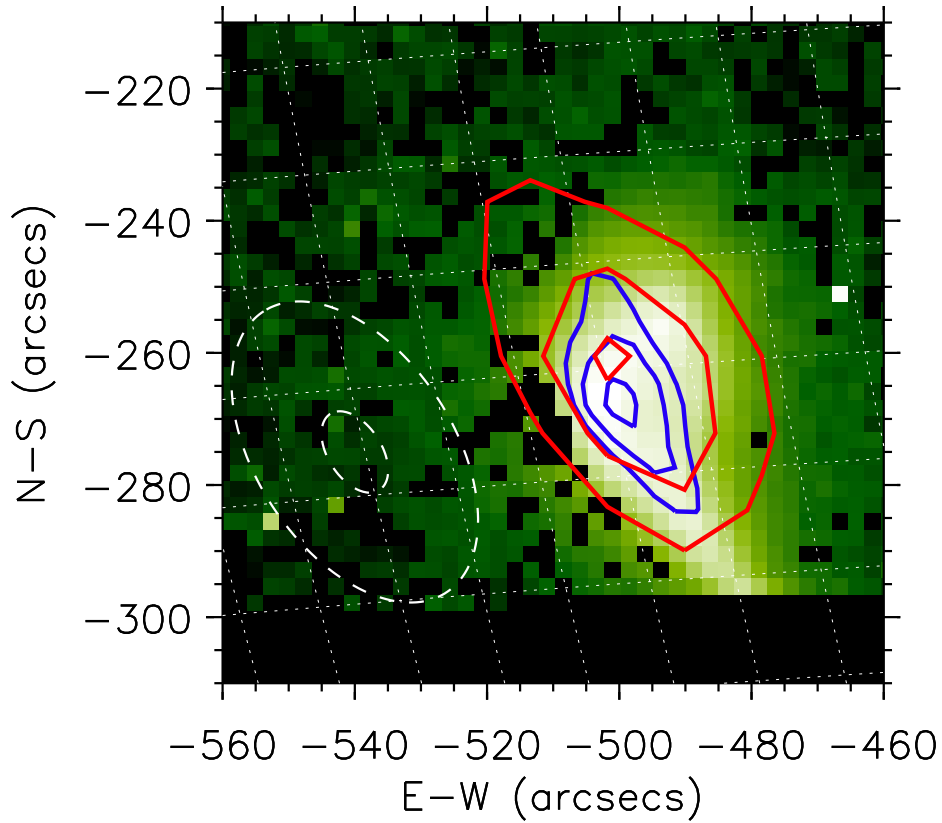}
\caption{2001 April 06. Emission measure (EM) map
(19:22:00 UT) derived from the Yohkoh SXT instrument \citep{c24},
using data obtained with two different filters (Be119 and Al12). For
clarity, only the OVSA (19:22:03 UT) RCP 2 GHz (red contours) and
LCP 7.4 GHz (blue contours) are overlaid here. The EM map reveals
the existence of a magnetic loop or arcade of loops filled with hot
and dense plasma, which is consistent with the magnetic and radio
topology presented in Fig. 2. The 2 GHz RCP radio source and the 7.4
GHz LCP kernel are well aligned with the most dense section of the
loop.}
\end{figure}

\clearpage

\begin{figure}
\plotone{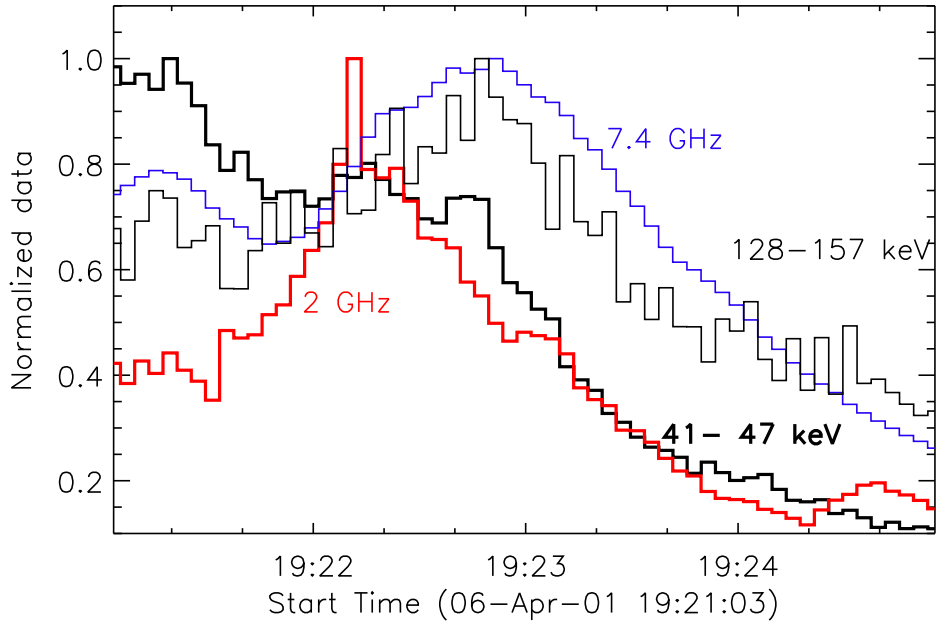}
\caption{OVSA total power lightcurves at 2 GHz (red
line) and 7.4 GHz (blue line), and Yohkoh \citep{c24} WBS  hard
X-ray counts in the 41-47 keV (thick line) and 128-157 keV (thin
line) ranges. Each curve has been normalized to the corresponding
maximum values recorded after 19:21 UT (3700 sfu at 2 GHz, 2300 sfu
at 7.4 GHz, 2088, and 244 HXR counts, respectively). The 128-157 keV
hard X-ray and 7.4 GHz time profiles are similar, which is
consistent with the 7.4 GHz emission being due to electrons of
energy $>300$ keV. The RTR emission at 2 GHz peaks about 1 min
earlier, has quite different time behavior, and best correlates with
the 41-47 keV hard X-rays channel, reflecting the fact that it is
due to a lower-energy part of the same population, and also depends
on other parameters such as the level of density fluctuations.}
\end{figure}

\end{document}